\documentclass[prl,amsmath,amssymb,showpacs,twocolumn,letterpaper]{revtex4}
\usepackage{graphicx}

\begin{document}

\title{Flux-tunable barrier in proximity Josephson junctions}

\author{Jian Wei$^1$}
\author{P. Cadden-Zimansky$^1$}
\altaffiliation[Present adress: ]{Department of Physics, Columbia University, New York, NY 10027, USA}
\author{P. Virtanen$^2$} 
\author{V. Chandrasekhar$^1$}
\affiliation{$^1$Department of Physics and Astronomy, Northwestern University, Evanston, IL 60208, USA\\
$^2$Low Temperature Laboratory, Helsinki University of Technology, Helsinki, Finland}

\date{\today}


\begin{abstract}
We report experiments on micron-scale normal metal loop connected by superconducting wires, where the sample geometry enables full modulation of the thermal activation barrier with applied magnetic flux, resembling a symmetric quantum interference device. We find that except a constant factor of five, the modulation of the barrier can be well fitted by the Ambegaokar-Halperin model for a resistively shunted junction, extended here to a proximity junction with flux-tunable coupling energy estimated using quasiclassical theory. This observation sheds light on the understanding of effect of thermal fluctuation in proximity junctions, while may also lead to an unprecedented level of control in quantum interference devices.
\end{abstract}
\maketitle

The dynamics of a particle trapped in a shallow potential well is a fundamental problem that has wide applicability to a number of areas in statistical physics.  At finite temperatures, due to coupling to a thermal bath, the particle may escape from the potential well through the process of thermal activation over the barrier represented by the edge of the potential.  At low temperatures, there may not be enough thermal energy for the particle to overcome the barrier, but the particle may retain a finite probability of escaping from the well  by tunneling through the barrier.  

For superconductors, two specific phenomena have been explored extensively in terms of the physics of thermal activation.  The first phenomenon is the generation of phase-slips in a thin superconducting wire with cross-section size comparable to the superconducting coherence length $\xi_S$~\cite{Little1967,Langer1967}.  In this case,  thermally activated phase slips (TAPS) lead to the appearance of a resistance tail of the thin wire at temperature below the nominal transition temperature $T_c$.  The voltage $V$ generated by TAPS is related to the time evolution of the macroscopic superconducting phase $\varphi$  through the Josephson relation $2eV/\hbar = d\varphi/dt$, which on average is determined by number of TAPS per unit time, each TAPS corresponding to a change of $2 \pi$.  At the instant in space and time where such a phase slip event occurs, the superconducting order parameter vanishes, which costs an energy $\Delta F$, the barrier over which the system must be thermally activated.

The second phenomenon is the onset of finite voltage in a Josephson junction~\cite{Ambegaokar1969}.  A resistively shunted junction (RSJ) can be modeled as a particle in a one-dimensional washboard potential in a viscous medium, where the distance coordinate corresponds to the phase difference $\varphi$ across the junction~\cite{Tinkham1996}.  With no current through the junction, the system sits in a local minimum of the potential.  Application of a current through the junction corresponds to tilting the washboard, shifting the position of the local minima.  At some value of current less than the nominal critical current $I_c$, the washboard potential is tilted sufficiently for the system to be thermally activated over the barrier between two adjacent potential minima.  Once this occurs, the system will continue to roll down the washboard potential, corresponding to a continuous time evolution of $\varphi$, and hence a finite voltage will appear across the junction according to the Josephson relation. 

The Josephson junction case leads to an interesting extension, where two junctions can be connected in parallel to form a dc superconducting quantum interference device (SQUID).  In this case, there are two independent parameters (the phase differences across the two junctions), giving rise to a two-dimensional potential for the system.  The system can transition from one local minimum to another through saddle-points in the potential~\cite{Tesche1977}. Due to the fact that the position of the system on the two dimensional potential is sensitive to the external magnetic flux $\Phi$,  dc SQUIDs have been investigated extensively due to their device potential~\cite{Clarke2004,Hopkins2005}. With any real dc SQUID, the two Josephson junctions cannot be fabricated to be exactly the same, which restricts the ability to tune the system.  Here we discuss doubly connected devices that are formed from normal metals in contact with superconductors.  The devices exhibit many of the properties of dc SQUIDs, but with the advantage that the devices can be designed to be almost perfectly symmetric, and hence allow unprecedented tunability of the system by means of an external magnetic flux.

\begin{figure}
\includegraphics[width=8.5cm]{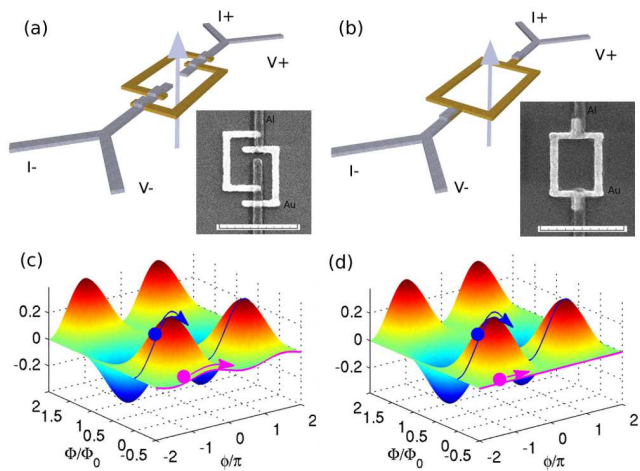}
\caption{(Color online) Schematic diagrams of two SNS quantum interference devices: asymmetric (a) and symmetric (b). The normal metal arms are shown in gold and the superconducting wires are shown in gray. The arrow in both figures corresponds to the the direction of applied magnetic flux. Inset: Scanning electron micrographs of the devices measured, the scale bar is 1 $\mu$m.  The calculated energy profile of the asymmetric/symmetric device are shown in (c)/(d) as a function of the external magnetic flux $\Phi$ and the phase difference across the two superconductors $\phi$, which is determined by the external current through the device. The energy profiles are calculated based on the quasiclassical theory, as described in the text. }  
\label{fig1}
\end{figure}

When a junction is made by a normal metal wire between two superconductors, such an SNS junction exhibits properties different from a conventional tunneling junction. The proximity to the superconductor has two major effects on the quasiparticles in the normal metal~\cite{Pannetier2000}.  First, it induces superconducting-like correlations between quasiparticles that increase the conductance of the normal metal, and second, it induces a gap in the quasiparticle density of states $N(E)$~\cite{leSueur2008}.  While in the superconductor the energy scale for the density of states is given by $\Delta$, the energy scale for $N(E)$ in the normal metal is set by the Thouless energy $E_{Th}= \hbar D/L^2$, where $D = v_F \ell /3$ is the quasiparticle diffusion in the normal metal, $\ell$ being the elastic scattering length, and $L$ is the length of the normal metal.   Similarly,  the maximal supercurrent that can flow through such a proximity junction is set by $E_{Th}$, not by $\Delta$ for a conventional Josephson tunnel junction, when $E_{Th} \ll \Delta$ (the long junction limit)~\cite{Wilhelm1997}. 

As with conventional junctions, two SNS junctions may be combined in parallel to form a dc SQUID.  Schematic of two types of SNS junctions are shown in Fig.~1. In the asymmetric device shown in Fig.~1(a), the normal metal arms (shown in gold) connect to the superconducting wires (shown in gray) at different points, resulting in likely different NS interface transparencies for the two arms of the device. Similar to a conventional dc SQUID, differences between the dimensions of the normal metal in the two arms, and differences between the NS interfaces in the two arms of the loop invariably lead to an asymmetric dc SQUID~\cite{Angers2008,Wei2008a}. However, the long-range nature of the Josephson coupling in SNS devices enables one to design and fabricate interference devices that are not possible with conventional tunnel junctions.  One such device is shown schematically in Fig.~1(b), where the interference device consists of a single normal metal loop between the two superconducting contacts, so that the NS interface transparencies for the two arms of the device are the same.  In this device, the modulation of quantum interference by an external magnetic flux occurs within the junction itself, i.e., the superconducting phase winding happens along the loop inside the junction, similar to that in a superconducting loop that shows classical Little-Parks oscillations~\cite{Little1962,Moshchalkov1993,Liu2001}.   As we shall see, because the NS interfaces are the same for both arms of the loop, this device behaves like a perfectly symmetric dc SQUID, while it is actually a single junction with a flux-tunable barrier.  From another point of view, it can also be described as a thin superconducting wire with a flux-tunable phase-slip center~\cite{note1}.

To understand the modulation of thermal activation barrier in these SNS junctions, we calculate the energy of the system as a function of the phase difference $\varphi$ between the two superconductors.  The energy of the system is given by~\cite{Tinkham1996b} 
\begin{equation}
  E_J (\varphi, \Phi) = \frac{\hbar}{2e} \int^\varphi d\varphi ' I_s(\varphi ', \Phi),
\label{eq1}
\end{equation}          
where $I_s(\varphi ', \Phi)$ is the supercurrent through the system, which is a periodic function of the phase difference $\varphi '$ and the externally applied flux $\Phi$.  To calculate the supercurrent, we use the extended circuit theory~\cite{Stoof1996} and numerically solve the Usadel equations for the sample geometries shown in Figs.~1(a) and 1(b). 
We assume here that the interface between the normal metal (N) and the superconductors (S) is perfectly transparent, the gap $\Delta$ regains its bulk value in the superconductor within a very short distance of the NS interface, and that the distribution of quasiparticles in the superconducting reservoirs is given by the equilibrium Fermi function $f(E)$. Since the characteristic unit for the supercurrent for an SNS junction is $E_{Th} /eR$~\cite{Wilhelm1997},  we use $(\hbar/2e^{2}R)E_{c}$ as the characteristic unit of energy for $E_J$. For the parameters used in this simulation, the amplitude of the supercurrent is about 0.2 $E_{Th} /eR$, so the  modulation of $E_J$ at fixed $\Phi$ is about 0.4 $(\hbar/2e^{2}R)E_{c}$ (note that since $E_J$ has an arbitrary constant from the integration in Eq.~(1), in Fig.~1 we assume $E_J =0$ at $\varphi =-2 \pi$). 

The resulting energy profiles for the system are shown in Figs.~1(c) and 1(d) as a function of $\varphi$ and $\Phi$, the two parameters under external experimental control.   If $\Phi$ is fixed at integral values of the superconducting flux quantum $\Phi_0=h/2e$, for both geometries, there is an energy barrier for evolution of the phase, as shown by the trajectories of the blue particles at $\Phi=0$, so that at low temperatures, the phase $\phi$ is stationary, and no voltage is developed across the device.
For half-integral values of the applied flux ($\Phi=(n+1/2)\Phi_0$, where $n$ is an integer) there is a difference between the asymmetric and symmetric cases.  For the asymmetric case, there is still an energy barrier, as shown by the trajectory of the red  particle in Fig.~1(c), although its height is smaller than that at integral values of the applied flux $\Phi=n\Phi_0$.  Consequently, although the resistance of the device at $\Phi=(n+1/2)\Phi_0$ will remain finite at first even when the resistance at $\Phi=n\Phi_0$ vanishes, the resistance at half-integral flux quanta will eventually vanish if the temperature is low enough, in the absence of quantum tunneling.  In contrast, for the perfectly symmetric case, there is no energy barrier at half-integral flux quanta (as shown by the trajectory of the red particle in Fig~1(d)), so that the device will have a finite resistance even at the lowest temperatures, since $\varphi$ can evolve in time without any energy cost.  

Such a simplified physical picture can well explain the magnetoresistance of the SNS junctions shown the insets of Fig.~1(a) and 1(b).  The details of fabrication and measurement are similar to those reported elsewhere~\cite{Wei2008a}. Figures 2(a) and 2(b) show the resistance of the asymmetric and symmetric devices respectively as a function of applied magnetic flux at a number of temperatures. 
The data are taken in the limit of zero dc current, with only a very small ac current for the resistance measurement (about 10-20 nA).  For both geometries, the resistance is finite at all values of $\Phi$ at higher temperatures, and is periodic in the applied flux, with a fundamental period of $\Phi_0$.  As the temperature is lowered, the resistance for both devices vanishes near $\Phi=n\Phi_0$.  As the temperature is lowered still further, the resistance of the asymmetric device also vanishes at half-integral values $\Phi=(n+1/2)\Phi_0$ of the applied flux.  In contrast, the resistance of the symmetric device at $\Phi=(n+1/2)\Phi_0$ remains finite down to the lowest temperatures.

\begin{figure}
\includegraphics[width=8.5cm]{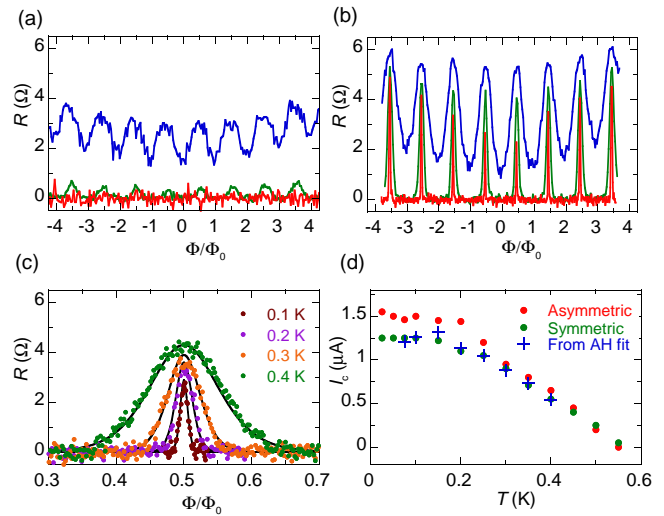}
\caption  {(Color online)  Resistance as a function of applied magnetic flux for the asymmetric sample (a) and the symmetric sample (b)  at 0.6  K (blue curves), 0.4 K (green curves), and 0.03 K (red curves).  While the oscillations for the asymmetric device die out rapidly with decreasing temperature, the oscillations for the symmetric device survive to the lowest measurement temperature.  (c)  Magnetoresistance of the symmetric device around $\Phi=\Phi_0/2$ at four different temperatures.  The solid lines are fits to the AH theory as described in the text.  Below 0.15 K, the magnetoresistance does not change with temperature.  (d)   Critical current at zero applied flux for the symmetric and asymmetric devices.  The plus symbols show the critical current expected from the fits of (c), multiplied by a factor of 5, as described in the text.} 
\label{fig2}
\end{figure}

The remarkable feature of the magnetoresistance curves for the symmetric sample is that peaks in resistance around $\Phi=(n+1/2)\Phi_0$ narrow as the temperature is lowered, saturating below about 0.15 K, as shown for the peak around $\Phi=\Phi_0/2$ in Fig.~2(c).  According to the physical picture in Fig.~1(d), there is a small but finite barrier for the system to overcome, except at exactly half-integral values of flux.  This barrier decreases monotonically as the system approaches $\Phi=(n+1/2)\Phi_0$.  Thus, as the temperature is lowered, the system needs to be closer to $\Phi=(n+1/2)\Phi_0$ so that the  particle can jump over the barrier and exhibit a finite resistance.  At low enough temperatures, the system may tunnel through the barrier, resulting in a temperature independent resistance at lower temperatures. 

To quantitatively model the magnetoresistance around half-integral values of the applied flux for the symmetric device, we use the thermal activation theory proposed by Ambegaokar and Halperin (AH) for a resistively shunted junction~\cite{Ambegaokar1969}, as extended to a dc-SQUID.  The AH theory predicts that the normalized resistance in the zero current limit is given by 
\begin{equation}
  R_{AH} = I_0^{-2} (\gamma/2)
\label{eq2}
\end{equation}
where $I_0$ is the modified Bessel function, and $\gamma=2E_j/k_B T$ is the ratio between the barrier and the thermal fluctuation energy.  In the standard RSJ model, the Josephson coupling energy is  $E_j = (\hbar/2e) I_c$, $I_c$ being the critical current.  In our SNS junction case, $E_j$ is given by Eq.~(\ref{eq1}), and is a function of the external flux $\Phi$.   Assuming symmetrical long SNS junctions, we obtain the usual sinusoidal dependence of the current $I_s$ on $\varphi$, and the Josephson coupling energy
\begin{equation}
E_j = \frac{\hbar I_c(\Phi)}{2e},
\label{eq3}
\end{equation}
where $I_c(\Phi) = I_c(0) |\cos (\pi\Phi/\Phi_0)|$ as for a symmetric dc SQUID~\cite{Tinkham1996c}.  Using this dependence for $I_c(\Phi)$, we have $\gamma = \hbar I_c(\Phi) /e k_B T$, and we can then fit the magnetoresistance of the symmetric device near $\Phi=\Phi_0/2$ using the AH theory with Eq.~(2).  

The solid lines in Fig.~2(c) show the resulting fits to the AH theory at four different temperatures, using only the measured peak resistance $R_p$ at $\Phi=\Phi_0/2$ and $I_c(0)$ as fitting parameters.  The resulting values of $I_c(0)$ are a factor of 5 smaller than the experimentally measured values of $I_c(0)$ over the entire temperature range.  Figure~2(d) shows a comparison of $I_c(0)$ obtained from the fits (multiplied by a factor of 5) compared to the experimentally measured values of $I_c(0)$. We note that the measured critical current saturates at lower temperatures and is about 10 times smaller than the value predicted at zero temperature limit~\cite{Wilhelm1997,Angers2008}, most probably due to an imperfect SN interface and heating~\cite{Courtois2008}. 

The strong  correlation between the measured and the fitted values indicates that the AH theory is applicable to diffusive SNS structures near $\Phi=\Phi_0/2$, but the effective potential barrier for thermal activation appears to be smaller by about a factor of 5 compared to Eq.~(\ref{eq3}). This discrepancy could not be due to quantum fluctuation~\cite{Liu2001} as this factor is temperature independent. It also can not be due to heating as then the measured critical current and the activation barrier should be smaller than that inferred from the AH fit at zero current limit.  In our opinion, this discrepancy could be due to the assumption of a standard external shunt resistor as the thermal noise source in the AH model, while the SNS junction is self-shunted. Another possibility is due to the neglect of inverse proximity effect and formation of minigap in both the normal metal and superconducting electrodes~\cite{leSueur2008}, which may change the relation between supercurrent and energy in Eq.~(\ref{eq3}). 

We note in previous measurements~\cite{Hoss2000}, the current voltage characteristics  of SNS junctions have shown large deviation from the AH theory and RSJ model, and theoretically it is not clear how thermal fluctuation effects the proximity junctions. Here, being able to tune the barrier at zero current limit (close to  equilibrium) enables us to apply the thermal activation theory  to SNS devices for the first time. This finding provides a first step for a coherent understanding of SNS devices and facilitates their applications such as sensitive detectors\cite{Giazotto2008}.

In summary, we have shown that diffusive SNS structures have the potential to provide an unprecedented level of control in quantum interference devices.  In particular, we have demonstrated that one can make an SNS junction with a flux-tunable barrier, which behaves like a perfectly symmetric dc SQUID. The potential barrier of such a device can be tuned to near zero with applied flux, something that is extremely difficult to achieve with conventional dc SQUIDs.       
            
This research was conducted with support from the National Science Foundation under grant No. DMR-0604601.  We thank V. Ambegaokar, P.M. Goldbart, A.A. Golubov, I.P. Nevirkovets, J.A. Sauls, S.E. Shafranjuk, H.H. Wang, F. Wilhelm,  
and A.D. Zaikin for helpful discussions.


\begin{thebibliography}{10}

\bibitem{Little1967} William~A. Little,  Phys. Rev. {\textbf 156}, 396 (1967).

\bibitem{Langer1967} J.~S. Langer and Vinay Ambegaokar, Phys. Rev. {\textbf 164}, 498 (1967);
D.~E. McCumber and B.~I. Halperin,  Phys. Rev. B {\textbf 1}, 1054 (1970).

\bibitem{Ambegaokar1969}Vinay Ambegaokar and B.~I. Halperin,  Phys. Rev. Lett. {\textbf 22}, 1364 (1969);
Y.~M. {Ivanchenko} and L.~A. {Zil'Berman}, JETP Lett. {\textbf 8}, 113 (1968).

\bibitem{Tinkham1996} See e.g., Michael Tinkham, {\em Introduction to Superconductivity},  McGraw Hill, 2nd edition, 1996, page 202.

\bibitem{Tesche1977}
C.~D. Tesche and J.~Clarke, J. Low Temp. Phys. {\textbf 29}, 301 (1977);C.~D. {Tesche}, J. Low Temp. Phys. {\textbf 44}, 119 (1981); 

\bibitem{Clarke2004}
J.~Clarke and A.I. Braginski,  {\em The SQUID handbook}, Wiley-VCH, 2004.

\bibitem{Hopkins2005}
David~S. Hopkins, David Pekker, Paul~M. Goldbart, and Alexey Bezryadin, Science {\textbf 308}, 1762 (2005).

\bibitem{Pannetier2000}
 B. Pannetier and H. Courtois, J. Low Temp. Phys. {\bf 188}, 599 (2000);
 H. Courtois, P. Charlat, Ph. Gandit, D. Mailly, and B. Pannetier, J. Low Temp. Phys. {\bf 116}, 187 (1999);
W. Belzig,  F.K. Wilhelm, C. Bruder, G. Sch\"on, and A.D. Zaikin, Superlatt. Microstruc. {\bf 25}, 1251 (1999).

\bibitem{leSueur2008}
H.~le~Sueur, P.~Joyez, H.~Pothier, C.~Urbina, and D.~Esteve, Phys. Rev. Lett. {\textbf 100}, 197002 (2008);
A.~K. Gupta, L.~Cretinon, N.~Moussy, B.~Pannetier, and H.~Courtois, Phys. Rev. B {\textbf 69}, 104514 (2004).

\bibitem{Wilhelm1997}
F.~K. {Wilhelm}, A.~D. {Zaikin}, and G.~{Sch{\"o}n},  J. Low Temp. Phys. {\bf 106}, 305 (1997);
 P. Dubos, H. Courtois, B. Pannetier, F.K. Wilhelm, A.D. Zaikin, and G. Sch\"on, Phys. Rev. B {\bf 63}, 064502 (2001);
J.~C. Hammer, J.~C. Cuevas, F.~S. Bergeret, and W.~Belzig, Phys. Rev. B {\bf 76}, 064514 (2007).

\bibitem{Angers2008}
L.~Angers, F.~Chiodi, G.~Montambaux, M.~Ferrier, S.~Gu\'{e}ron, H.~Bouchiat,
  and J.~C. Cuevas, Phys. Rev. B {\bf 77}, 165408 (2008).

\bibitem{Wei2008a}
J.~Wei, P.~Cadden-Zimansky, and V.~Chandrasekhar, Appl. Phys. Lett. {\bf 92}, 102502 (2008).

\bibitem{Little1962}
W.~A. Little and R.~D. Parks, Phys. Rev. Lett. {\bf 9}, 9 (1962); Ref.[4], page 128.

\bibitem{Moshchalkov1993}
V.~V. {Moshchalkov}, L.~{Gielen}, M.~{Dhalle}, C.~{van Haesendonck}, and
  Y.~{Bruynseraede},  Nature {\bf 361}, 617 (1993).

\bibitem{Liu2001}
Y.~Liu, Yu. Zadorozhny, M.~M. Rosario, B.~Y. Rock, P.~T. Carrigan, and H.~Wang, Science {\bf 294}, 2332 (2001).

\bibitem{note1}
This point is demonstrated by the temperature dependence of the resistance as well as the current voltage characteristics, which will be presented elsewhere in more details.

\bibitem{Tinkham1996b}
See e.g., ref.[4], page 198.

\bibitem{Stoof1996}
T.~H. Stoof and Yu.~V. Nazarov, Phys. Rev. B {\bf 54}, R772 (1996);
A.~A. Golubov, F.~K. Wilhelm, and A.~D. Zaikin, Phys. Rev. B {\bf 55}, 1123 (1997).

\bibitem{Tinkham1996c} See e.g., ref.[4], page 215.

\bibitem{Courtois2008}
H.~Courtois, M. Meschke, J.~T. Peltonen, and J.~P. Pekola, Phys. Rev. Lett. {\bf 101}. 067002 (2008).

\bibitem{Hoss2000}
T.~Hoss, C.~Strunk, T.~Nussbaumer, R.~Huber, U.~Staufer, and
  C.~Sch\"onenberger, Phys. Rev. B {\bf 62}, 4079 (2000); 
P.~Dubos, H.~Courtois, O.~Buisson, and B.~Pannetier, Phys. Rev. Lett. {\bf 87}, 206801 (2001);
Tatsushi Akazaki, Hayato Nakano, Junsaku Nitta, and Hideaki Takayanagi, Appl. Phys. Lett. {\bf 86}, 132505 (2005);
E.~Lhotel, O.~Coupiac, F.~Lefloch, H.~Courtois, and M.~Sanquer, Phys. Rev. Lett. {\bf 99}, 117002 (2007).


\bibitem{Giazotto2008}
Francesco Giazotto, Tero~T. Heikkil\"{a}, Giovanni~Piero Pepe, Panu
  Helist\"{o}, Arttu Luukanen, and Jukka~P. Pekola, Appl. Phys. Lett. {\bf 92}, 162507 (2008);
Francesco Giazotto, Tero~T. Heikkila, Arttu Luukanen, Alexander~M. Savin, and
  Jukka~P. Pekola, Rev. Mod. Phys. {\bf 78}, 217 (2006).

\end{thebibliography}
\end{document}